\documentclass[prc,twocolumn,showpacs,amsmath,amssymb]{revtex4}
\usepackage{times}
\usepackage{graphicx}
\usepackage{dcolumn}
\usepackage{bm}
\usepackage{amstext}

\begin{document}

\title{Structure effect on one neutron removal reaction using relativistic mean field densities in
 Glauber model}
\author{R. N. Panda$^1$, and S. K. Patra$^2$}
\affiliation{
$^1$Department of Physics, ITER, Siksha 'O' Anusandhan University,
Bhubaneswar-751 030, India, \\ 
$^2$Institute of Physics, Sachivalaya Marg, Bhubaneswar-751 005, India. 
}
\date{\today}

\begin{abstract}

We calculate the one neutron removal reaction cross-section  $(\sigma_{-1n})$
for few stable and neutron-rich halo nuclei with $^{12}$C as target, using 
relativistic mean field (RMF) densities, in the frame work of Glauber model.
The results are compared with the experimental data. Study of the stable
nuclei with the deformed densities have shown a good agreement with the data, 
however, it differs significantly for the halo nuclei. We observe that while
estimating the $\sigma_{-1n}$value from the difference of reaction cross-section 
of two neighboring nuclei 
with mass number A and that of A-1 in an isotopic chain, we get good agreement 
with the known experimental data for the halo cases.

\end{abstract}

\maketitle

\section{Introduction}

From last two decades or more, the exploration of neutron-rich nuclei is an 
important branch in Nuclear Physics research. It is a source of observance of 
new phenomena and dynamics. This is possible due to the development of accelerator techniques
for Radioactive Ion Beams (RIBs) in various laboratories around the globe. 
Experimental methods and theoretical analysis have been widely used to collect 
information about the structure information, such as nuclear size, valence nucleon distribution and halo 
structure of these exotic nuclei. The measurement of  various reaction observables 
like total reaction cross-section $\sigma_r$, one- and two- nucleon removal cross-section 
($\sigma_{-1n}$, $\sigma_{-2n}$) and the longitudinal momentum distribution $P_{||}$ are some of 
the established quantities for such studies.

The relativistic mean field (RMF) or the effective field theory motivated RMF, i.e.,
the E-RMF provides the internal structure  or sub-structure information of
the nuclei through the density distributions, using as an input
while calculating the observables in conjunction with Glauber model 
\cite{skpatra1,rnpanda1}. A systematic study of one- and two-neutron knockout 
data for $^{15-19}$C, explained beautifully while using shell model, which gives a 
consistent structure information not only for the stable nuclei but also neutrons at 
the boundary \cite{sim09}. It is well known that narrow fragment momentum 
distribution reflects large space distribution of the valence  nucleon and 
there is a correlation of the magnitude of the $\sigma_{-1n}$ with the width of 
the $P_{||}$, in approaching the nucleon (neutron or proton)  drip-lines. However, 
one-neutron removal reaction cross-sections provide important nuclear 
structure information complementary to that obtained from $P_{||}$.

In the present paper, our aim is to calculate the $\sigma_r$  
and $\sigma_{-1n}$ by 
using densities obtained from the RMF and E-RMF formalisms \cite{skpatra1,
rnpanda1} in conjunction with Glauber model. It will be shown for the nuclei near 
the drip-line which possess a halo-structure fails to be explained by the 
standard evaluation of $\sigma_{-1n}$. Contrary to this estimation, the 
difference in total reaction cross-section between two consecutive neighboring 
nuclei in an isotopic chain better matches with the experimental data.

\section{Theoretical Framework}

The use of RMF and E-RMF formalisms for finite nuclei as well as infinite
nuclear matter are well documented and details can be found in \cite{patra91,
lala97} and \cite{tang97, patra01} respectively.  
The working expressions for density profiles and other related quantities are available 
in \cite{skpatra1,rnpanda1,patra91,tang97,patra01}.
The details to calculate $\sigma_r$ using Glauber approach has been given
by R. J. Glauber \cite{gla59}. This model is based on the independent,
individual nucleon-nucleon ($NN$) collisions
in the overlapping zone of the colliding nuclei, and has been used extensively to
explain the observed total nuclear reaction
cross-sections for various systems at high energies. The standard Glauber form
for the total reaction cross-sections at
high energies is expressed as \cite{gla59,kar75}:
\begin{equation}
\sigma _{r}=2\pi\int\limits_{0}^{\infty }b[1-T(b)]db \;,
\end{equation}
where $T(b)$ is the transparency function with impact parameter $b$.
The function $T(b)$ is calculated in the overlap region between the projectile and the target assuming
the interaction is formed from a single $NN$ collision. It is given by
\begin{equation}
T(b)=\exp \left[ -\sum\limits_{i,j}\overline{\sigma }_{ij}\int d%
\vec{s}\overline{\rho }_{ti}\left( s\right) \overline{\rho }%
_{pj}\left( \left| \vec{b}-\vec{s}\right| s\right)
\right] \;.
\end{equation}
The summation indices $i$ and $j$ run over proton and nucleon and subscripts $p$ and $t$ referred to 
projectile and target, respectively. The experimental nucleon-nucleon reaction  cross-section 
$\overline{\sigma }_{ij}$ varies with energy. The $z$-integrated densities $\overline{\rho }(\omega )$ 
are defined as
\begin{equation}
\overline{\rho }(\omega )=\int\limits_{-\infty }^{\infty }\rho \left( \sqrt{%
\omega ^{2}+z^{2}}\right) dz \;,
\end{equation}
with $\omega ^{2}=x^{2}+y^{2}$.  The argument of $T(b)$ in Eq. (2) is $\left| \vec{b}-\vec{s}\right|$, which stands for
the impact parameter between the $i^{th}$ and $j^{th}$ nucleons.

The original Glauber model was designed for high energy approximation. However, it was found to work 
reasonably well for both the nucleus-nucleus reaction and the differential elastic 
scattering cross-sections over a broad energy range \cite{chau83}. To include 
the low energy effects of $NN$ interaction, the  Glauber model is modified to take care of the 
finite range effects in the profile function and Coulomb modified trajectories\cite{pshukla03,abu03}. 
The modified
$T(b)$ is given by \cite{abu03,bhagwat},
\begin{small}
\begin{equation}
T(b)=\exp
\left[-\int_{p}\int_{t}\sum
\limits_{i,j}\left[\Gamma _{ij} \left( \vec{b} - \vec{s} + \vec{t}
\right) \right] \overline{\rho}_{pi} \left( \vec{t} \right)
\overline{\rho }_{tj} \left( \vec{s}\right) d\vec{s}d\vec{t} \right].
\label{eq:2}
\end{equation}
\end{small}
          
\noindent The profile function $\Gamma_{ij}(b_{eff})$ is defined as \cite{skpatra1}

\begin{equation}
\Gamma _{ij}(b_{eff})=\frac{1-i\alpha _{NN} }{2\pi \beta _{NN}^{2}}\sigma
_{ij}\exp \left( -\frac{b_{eff}^{2}}{2\beta _{NN}^{2}}\right)\;,
\label{eq:3}
\end{equation}
with $b_{eff}=\left| \vec{b}-\vec{s}+\vec{t%
}\right| $,
$\vec{s}$ and $\vec{t}$ are the dummy variables for integration over
the $z$-integrated target and projectile densities. The parameters $\sigma _{NN}$, $\alpha _{NN}$, and $\beta _{NN}$ are usually
case-dependent (proton-proton, enutron-neutron or proton-neutron), but we have used the appropriate average values
from Refs. \cite{kar75,charagi90}.
The deformed or spherical nuclear densities obtained from the RMF and E-RMF models are fitted to a 
sum of two Gaussian functions with
suitable co-efficients $c_i$ and ranges $a_i$ chosen for the respective nuclei which is expressed as
\begin{equation}
\rho (r)=\sum\limits_{i=1}^{2}c_{i}exp[-a_{i}r^{2}].
\end{equation}
Then, the Glauber model is used to calculate the total reaction cross-section for both the stable 
and unstable nuclei considered in the present study.

The expression for one nucleon removal reaction cross-section $\sigma_{-1n}(I)$ is given by \cite{abu03}
\begin{equation}
\sigma_{-1n}(I)=\sum_{c}\int d\overrightarrow{k}\sigma_{a=(k,g=0),c}, 
\end{equation}
\noindent where $\sigma_{a=(k,g=0),c}$ are the possible final states $ac$.  
In the present formalism, it is considered that the projectile nucleus breaks up into a 
core and the removed nucleon. The core $C$ has an internal wave function $\phi_g$ and the 
one-nucleon, i.e.,
the departed nucleon has an asymptotic momentum $\hbar {\bf k}$ in the continuum state with respect to
the core.  The core is considered to be in the ground state ($g=0$) at the time of the
collision. The total $\sigma_{-1n}(I)$ can be separated to an elastic ($\sigma_{-1n}^{el}$) with $c=0$
and inelastic ($\sigma_{-1n}^{iel}$) part having $c$ as non-zero. The $\sigma_{-1n}^{el}$ 
and $\sigma_{-1n}^{iel}$ is expressed as \cite{abu03}
\begin{eqnarray}
\sigma_{-1n}^{el}(I)&=&\int d{\bf b}\{<\phi_0\mid e^{-2Im\chi_{Ct}(b_C) 
-2Im\chi_{-1nt}(b_C+s)}\mid \phi_0> \nonumber \\
&&-\mid <\phi_0\mid e^{-i\chi_{Ct}(b_C)
+i\chi_{-1nt}(b_C+s)}\mid \phi_0>\mid^2\} 
\end{eqnarray}

\begin{eqnarray}
\sigma_{-1n}^{iel}(I)&=&\int d{\bf b}\{<\phi_0\mid e^{-2Im\chi_{Ct}(b_C)} \nonumber \\
&&-e^{-2Im\chi_{Ct}(b_C)-2Im\chi_{-1nt}(b_C+s)}
\mid \phi_0>\},
\end{eqnarray}
\noindent here $\chi_{pt}$ is the phase shift function and $\phi_0$ is the valence
wave function (the wavefunction of the removed nucleon). The notation  
and the numerical procedure of calculation of one-nucleon removal reaction cross-section 
are followed from Ref. \cite{abu03}.  

\section{Calculations and Results}

We obtain the field equations for 
nucleons and mesons from the RMF and E-RMF lagrangian. 
For deformed case (RMF only), these equations are solved by expanding the upper and
lower components of the Dirac spinners and the boson fields in an axially
deformed harmonic oscillator basis. The set of coupled equations are solved
numerically by a self-consistent iteration method taking
different inputs of the initial deformation $\beta_0$
\cite{gam90,patra91}. For spherical densities both for RMF and
E-RMF models, we follow the numerical procedure of Refs. \cite{patra01}. 
In our calculation, the constant gap BCS pairing is used to add the pairing effects
for open shell nuclei. The centre-of-mass motion (c.m.) energy correction is estimated by the usual
harmonic oscillator formula $E_{c.m.}=\frac{3}{4}(41A^{-1/3})$.

\begin{figure}
\vspace{0.4cm}
\hspace{-0.3cm}
\includegraphics[scale=0.32]{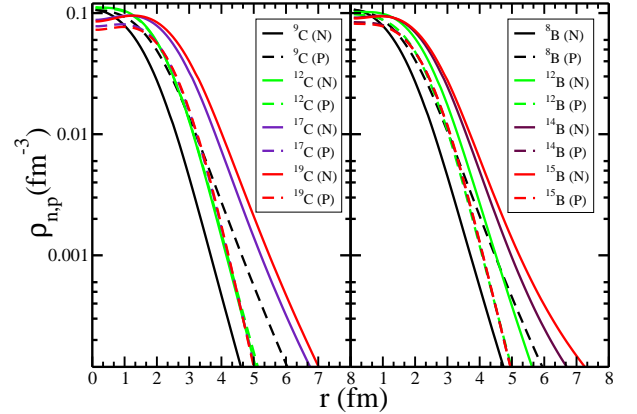}
\caption{\label{fig:epsart1}The spherical proton ($\rho_p$) and neutron
($\rho_n$) density obtained from RMF (NL3) parameter set for various
isotopes of (a) Carbon and (b) Boron.
}
\end{figure}

Comparing the binding energy (BE) of the calculated
solutions, the  maximum BE  and the corresponding densities
[$\rho_p$ (proton) and $\rho_n$ (neutron)] are for the ground state.
All other solutions are the excited intrinsic state including the spherical one.
Since the main input in the Glauber model estimation is the RMF or E-RMF densities, it is
important to have an information of these quantities. We have
plotted the spherical $\rho_p$ and $\rho_n$ for both proton and neutron of Carbon and
Boron isotopes in Figure 1 using RMF (NL3) parameter set \cite{lala97}. As expected, we 
find an extended density distribution for proton compared to neutron in case of $^{9}$C 
and $^{8}$B due to the proton-rich nature of these two nuclei. The value of $\rho_n$ and 
$\rho_p$ are almost similar for $^{12}$C which can be seen from Figure 1.  Extension of
$\rho_n$ is much more than $\rho_p$ for rest of the nuclei.  It is maximum for $^{19}$C 
and $^{15}$B in Carbon and Boron isotopic chains, respectively, because of the high neutron 
to proton ratio for these cases. 

\begin{table*}
\caption{\label{tab:table2}{One-neutron- for $^{12,13,15,17,19}C$ and one-proton removal reaction
cross-sections $\sigma_{-1n}$ (in mb) for $^9C$ and $^{8}$B with  $^{12}C-$target
using spherical RMF and E-RMF densities obtained from NL3 and G2 parameter
sets, respectively. The available experimental data
\cite{cort01,sau04,cuie05,ols83} is given for the comparison. The last occupied
proton and neutron single-particle energies are also given. 
}}
\bigskip
\begin{tabular}{|c|c|c|c|c|c|c|c|c|c|c|c|c|c|c|c|c|c|}
\hline\hline
Projectile& Energy& $\sigma_{-1n}$&\multicolumn{2}{c|}{$\sigma_{-1n}(I)$} &
\multicolumn{2}{c|}{$\sigma_{-1n}(II)$}&\multicolumn{2}{c|}{$\epsilon_{p} (MeV)$}
&\multicolumn{2}{c|}{$\epsilon_{n} (MeV)$} \\
& & Exp. & RMF& E-RMF & RMF & E-RMF&  RMF& E-RMF& RMF& E-RMF \\
\hline
$^{9}C$ & 285 &48(8)&81&96 &34&31&-4.07&-4.15&-13.81&-13.90 \\
$^{12}C$ & 1050 &44.7(3) &39&37 &55&39&-15.66&-13.16&-18.97&-16.16 \\
$^{13}C$ & 800 &&29&28&47&37&-16.81&-15.30&-17.77&-16.35 \\
$^{15}C$ & 54 &137(16) &95&88 &37&38&-19.18&-20.47&-10.57&-10.39\\
$^{17}C$ & 904 &129(22) &101&95 &45&32&-23.93&-22.65&-11.53&-11.57\\
$^{19}C$ & 910 &231(51) &134&128 &35&37&-27.25&-25.89&-12.61&-12.50\\
$^{8}B$ & 285 &89(2)&103&107&47&33&-23.16&-20.73&-31.50&-28.96 \\
$^{12}B$ &67  &81(5) &54&49&47&44&-39.95&-34.18&-14.20&-13.49\\
$^{13}B$ & 57 &59(4)&54&38&46&43&-39.88&-35.49&-7.81&-7.85 \\
$^{14}B$ & 50 &153(15)&70&64&48&38&-41.00&-36.85&-8.40&-8.41 \\
$^{15}B$ & 43 &108(13)&87&89&51&39&-42.16&-38.23&-8.77&-8.96 \\
\hline\hline
\end{tabular}
\label{Table 1}
\end{table*}

In the present study of $\sigma_r$ and $\sigma_{-1n}$, first we use the
spherical density obtained from RMF (NL3)\cite{lala97}  and E-RMF (G2) \cite{tang97}.
The results are presented in Table 1 for $^{9,12,13,15,17,19}C$ and $^{8,12,13,14,15}B$ 
isotopes with $^{12}C-$target at various projectile energies. These results deviate considerably 
from the data \cite{cort01,sau04,cuie05,ols83} which are quoted in the table. For example, in case
of $^9$C+$^{12}$C, the observed value of $\sigma_{-1n}$ is $48\pm8$ mb as compared to the estimated 
results of 81 and 96 mb with NL3 and G2 parametrization, respectively. Note that the
$\sigma_{1n}$ for $^8B+^{12}C$ and $^9C+^12C$ systems are one-proton removal
reaction cross-section, which may be followed throughout the text and Tables. However,
in rest of the systems, $\sigma_{1n}$ will be refered as one-neutron removal reaction cross-section.  
Similar discrepancy is also 
seen for other cases. A further inspection of the table shows that the experimental one-neutron
removal reaction cross-section for some selected cases coincide well with the prediction.
We also used the method of B. Abu-Ibrahim et al. \cite{abu03} to calculate the one-neutron removal 
reaction cross-section $\sigma_{-1n}(II)$, which obtain by the difference of total reaction 
cross-section of two neighboring nuclei with mass number A and A-1 in an isotopic chain. This
prescription is suitable only for halo projectile and may not be applicable for general cases.
The expression is given by \cite{abu03}:
\begin{equation}
\sigma_{-1n}(II)=\sigma_{r}(^{A}Z)-\sigma_{r}(^{A-1}Z),
\end {equation}
and the values are inserted in Table 1 for comparison. The $\sigma_{-1n}(II)$ differs significantly 
from the experimental data for all the cases.  It is 
important to recall that the effect of deformation is nominal in the evaluation of $\sigma_r$ which 
is reported in our earlier publications \cite{skpatra1,rnpanda1}.  In these papers, the Glauber 
model with RMF (NL3, NL-SH) and E-RMF (G2) densities show a good agreement with experimental 
data for $\sigma_r$ and elastic differential scattering cross-sections $d\sigma/d\Omega$, which
in general justify the model independency of the calculation with various relativistic parametrizations. 

\begin{table*}
\caption{\label{tab:table1}{Same as Table 1, but using both spherical and deformed NL-SH densities
for $^{9,12,15,17,19}C$ and $^{8,12,13,14,15}B$ projectiles
taking  $^{12}C$ as target. The available
experimental data is also displayed for the comparison \cite{cort01,sau04,cuie05,ols83}.
The last occupied neutron and proton single-particle orbits are also given and it is
denoted by the Nilsson index $[Nn_3\Lambda]\Omega^{\pi}$.}
}
\begin{tabular}{|c|c|c|c|c|c|c|c|c|c|c|c|c|c|c|c|c|}
\hline\hline
Projectile&Target& Energy& $\sigma_{-1n}$&\multicolumn{2}{c|}{$\sigma_{-1n} (I)$} &
\multicolumn{2}{c|}{$\sigma_{-1n}(II)$}& $\beta_{2}$ &\multicolumn{2}{c|}{neutron}&\multicolumn{2}
{c|}{proton}\\
&&&Exp.&Sph.& Def.&Sph. & Def.& & $[Nn_3\Lambda]$&$\epsilon_n (MeV)$
& $[Nn_3\Lambda]$&$\epsilon_p (MeV) $\\
\hline
$^{9}C$ &$^{12}C$& 285& 48(8)&70&66 &24&03&0.36
&$[110]1^-$& -16.122&$[101]3^-$&$-2.772$ \\
$^{12}C$ &$^{12}C$& 1050 &44.7(3) &39&49 &45&293&-0.21
&$[101]1^-$& -16.952&$[101]1^-$&$-13.752$ \\
$^{15}C$ &$^{12}C$ &54& 137(16)&92&130&35&15&0.25
&$[220]1^+$&$-2.611$&$[101]3^-$&-19.631\\
$^{17}C$ &$^{12}C$ &904& 129(22)&122&120&31&32&0.45
&$[211]3^+$&$-2.705$&$[101]3^-$&-21.906\\
$^{19}C$ &$^{12}C$ &910& 233(51)&134&152&31&263&-0.43
&$[202]3^+$& $-3.316$&$[101]1^-$&$-27.048 $\\
$^{8}B$ &$^{12}C$ &285 &89(2)&101&99 &34&10&0.63
&$[110]1^-$ &-14.054&$[101]3^-$&$-1.880 $\\
$^{12}B$ &$^{12}C$ &67 &81(5)&41&65 &45&70&0.18
&$[101]1^-$ &-6.212&$[101]3^-$&$-15.585 $\\
$^{13}B$ &$^{12}C$ &57 &59(4)&53&46 &44&16&0.10
&$[110]1^-$& -7.378&$[101]3^-$&$-18.321 $\\
$^{14}B$ &$^{12}C$ &50 &153(15)&67&97 &32&15&0.38
&$[220]1^+$ &-1.992&$[101]3^-$&$-18.686 $\\
$^{15}B$ &$^{12}C$ &43 &108(13)&88&104 &68&73&0.59
&$[220]1^+$ &-2.611&$[101]3^-$&$-19.631 $\\
\hline\hline
\end{tabular}
\label{Table 2}
\end{table*}
Unlike to the total reaction cross-section, the $\sigma_{-1n}$ obtained
from the Glauber model, depends very much on the structure information 
of the projectile and target nuclei, i.e., input densities of these systems. 

The nuclear single-particle energy $\epsilon_{n,p}$ for the last occupied orbit is very
important for a reaction process. Thus, it is worthwhile to analyse the  $\epsilon_{n,p}$
of the valence nucleon of the projectile and target nuclei. For simplicity, the spherical 
single-particle energy for the last occupied orbit for proton $\epsilon_p$ and neutron 
$\epsilon_n$ with RMF (NL3) and E-RMF (G2) are compared. As expected, a small variant in
$\epsilon_p$ or $\epsilon_n$ makes a remarkable change in $\sigma_{-1n}(II)$ [Eq. (10)] for
many cases. For example, the one neutron removal reaction cross-section $\sigma_{-1n}(II)$,
for $^{12}C+^{12}C$ are 55 and 39 mb for RMF (NL3) and E-RMF (G2) with
their single-particle energies $\epsilon_p=-15.66$, $\epsilon_n=-18.97$ and
$\epsilon_p=-13.16$, $\epsilon_n=-16.16$ MeV, respectively. This discrepancy is minimum in
the calculation of $\sigma_{-1n}(I)$ [Eq. (7)] with a lone exception for
$^{13}B+^{12}C$ system.
As the valence $\epsilon_{n,p}$ plays a major role to determine the reaction observables,
one needs to reproduce these values with the experimental observation. This can be achieved by a
small adjustment of the parameters in the relativistic mean field formalisms. However, the 
philosophy of RMF or E-RMF of single set of parametrization for the entire domain of nuclear landscape 
goes against this parameter fiddling.  Keeping this in mind, the quality 
of the results is compromised slightly using with the original values of NL3, NL-SH or G2 sets.

Apart from the single-particle energy, the structure effect of the participating
nuclei is crucial for a reaction study. In this context, it is interesting enough 
to see this effect (deformation effect) on $\sigma_{-1n}$. We repeat the
calculations for $\sigma_{-1n}(I)$ and $\sigma_{-1n}(II)$ with the deformed
densities (RMF only) as input in the Glauber model \cite{skpatra1}.
We obtain spherical equivalent of the axially deformed densities
using equations (3) and (6) following the prescription of Refs.
\cite{skpatra1,rnpanda1}. The NL-SH parameter set \cite{sharma93} for
this purpose is used and the results are listed in Table 2. The reason to change the NL3
to NL-SH is the unavailability of converged ground state deformed
solution with NL3 for very light mass nuclei, which is a situation in the present study
\cite{reinhard88}. Also, the NL-SH parametrization is reasonably a better parameter set
and we expect similar outcome from this force parameter.

Due to similar reason as mentioned for the spherical nuclei, the deformed densities for some selected cases
are imperative to analyse. Our earlier work on density study supply us enough signature about the
complicated sub-structure \cite{aru05}. The clustering and sub-structure of these deformed 
neutron and proton density distributions are demonstrated in Figure 2.  The density
contours presented are in boxes of width and height 6 fm. A uniform contour
spacing of 0.01 fm$^{-3}$ is used for proton and neutron densities.
The z-axis is chosen as the symmetry axis, the densities are evaluated in the $z\rho$ plane, 
where $x = y = \rho$.
In ref. \cite{aru05}, it is noticed that $^{12}$C possesses a $3\alpha-$cluster with a tetrahedral
configuration. The same structure is reproduced in the present study with an oblate shape. The
structure of the neutron deficient $^9$C nucleus has a prolate ground state and that of the
neutron-rich $^{19}$C has an oblate ground state deformation.  In all the three cases, the
density plot informs that
the central part of the nucleus is a compact core, which is surrounded by a thin layer
of nucleons. The structure of the internal core for both proton and neutron have different
density distribution from $^{9}$C to $^{19}$C. The shape of $^{19}$C proton density distribution
looks like a perfect dumb-bell. Thus, it has a maximum probability to
to have the structural effects on neutron
removal reaction. On the other hand, the total nuclear reaction cross-section is less
influenced by deformation, may be  because of the averaging in input density in the Glauber model 
calculations.

The results obtained from the deformed densities are tabulated (see Table 2).
Table 2 shows that most of the $\sigma_{-1n}(I)$ [obtained from Eq. (7)] 
matches quite well with the experimental data of \cite{cort01,sau04,cuie05,ols83} and
few of them do not agree. On the other hand $\sigma_{-1n}$ evaluated from Eq. (10) coincide
with only $^{19}C+^{12}C$ experimental data \cite{cort01}.
Of particular interest amongst the nuclei investigated here are $^{14}B$ and
$^{19}C$ which based on the relatively weak binding of the valence neutrons
and measurements of one neutron removal cross-sections have suggested
to be one neutron halo systems \cite{bazin98}. The single-particle energy for
proton $\epsilon_p$ and neutron $\epsilon_n$ for the last occupied orbit are given
in the 11th and 13th column of Table 2. The last proton for $^8B$ and $^{9}$C and
the outer most neutron for $^{14,15}B$ and $^{17,19}C$ are loosely bound which are the possible candidates
for either proton-halo (or skin) or neutron-halo (skin). Going back to the analysis of 
Figure 2, which gives us enough indication for the absence of halo-like structure in $^{9,12}$C.
Contrary to the case of $^9C$ and $^{12}C$ a thin-layer of neutron distribution spread spatially 
to a large extent in case of $^{19}C$, which
looks like a halo-nucleus. This behavior is also reflected in the one neutron removal
reaction cross-section. In this particular case of $^{19}$C, the calculated result $\sigma_{-1n}(II)$
= 263 mb is
more closure to the experimental value of $\sigma_{-1n}=233 \pm 51$ mb than the $\sigma_{-1n}$ obtained by
using Eqn. (7). If we
recall the statement of Abu-Ibrahim et al. \cite{abu03}, then $^{19}C$ is a halo-nucleus. On the
other hand, the measured one neutron removal reaction cross-section of  $^{14}B$ is larger
than its neighbors suggest the weak binding of the last neutron and extended
valence density distribution. 

Summarising the whole discussions of Tables 1 and 2, in general, one can say it very briefly that
except few cases like $^{12,17}$C+$^{12}$C
and $^{8,13}$B+$^{12}$C the spherical density used from RMF (NL3) and E-RMF (G2) fails to
reproduce the data. When we use the deformed densities to evaluate the one neutron
cross-section, the predicted $\sigma_{-1n}(I)$ matches reasonably well
with the experimental measurement. In this case, only the result of the system
$^{19}$C+$^{12}$C deviate from the observation. However, the results predicted
by Eq. (10) disagree largely with the experiments irrespective of the densities used. Complementating
to Eq. (7), the $\sigma_{1n}(II)$ matches with the lone case $^{19}$C+$^{12}$C agreeing
with the prediction of Ref. \cite{abu03}.

\begin{figure}
\vspace{0.4cm}
\hspace{-0.3cm}
\includegraphics[scale=0.32]{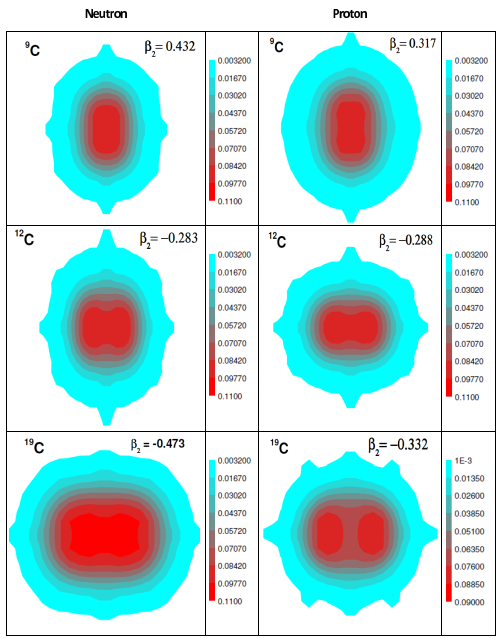}
\caption{\label{fig:epsart2}The axially deformed density distribution for $^{9,12,19}$C with
RMF (NL-SH) parameter set.
}
\end{figure}

\begin{figure}
\vspace{0.4cm}
\hspace{-0.3cm}
\includegraphics[scale=0.32]{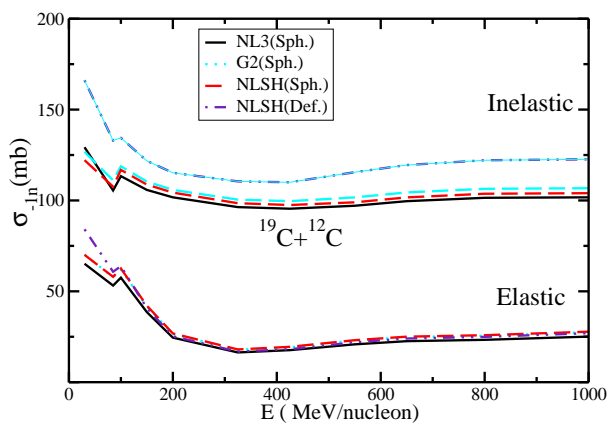}
\caption{\label{fig:epsart3} The energy dependence of the neutron removal cross-section for
$^{19}C$+ $^{12}C$ system using spherical densities of RMF (NL3), RMF (NL-SH)
and E-RMF (G2) parameter sets for both elastic
and inelastic processes. The result obtained by deformed RMF (NL-SH) densities is
also given for the comparison.
}
\end{figure}

In Figure 3, we have presented the  $\sigma_{-1n}(I)$ with various incident energies
for $^{19}C+^{12}C$ using the spherical NL3, NL-SH and G2 densities
in the Glauber model calculation. We also compare our results with the deformed NL-SH densities
obtained from the axially deformed RMF. All the spherical densities
reproduce similar elastic $(\sigma_{-1n}^{el})$ and inelastic $(\sigma_{-1n}^{iel})$
one neutron removal reaction cross-section. The deformed NL-SH densities has a large impact
on the evaluation of $\sigma_{-1n}$ unlike to the total nuclear reaction cross-section
$\sigma_r$, which is evident from Figure 3 and consistent with 
$\sigma_{-1n}(I)$. The deformed $\sigma_{-1n}^{iel}$ value constantly over-estimate the spherical
$\sigma_{-1n}^{iel}$ starting from low to very high incident energy of the projectile.

\section{ Summary and Conclusion}
In summary, one neutron removal reaction cross-sections for the neutron-rich isotopes
have been calculated using the densities obtained from RMF (NL3) and E-RMF (G2) formalisms
for spherical and deformed NL-SH parameter sets.  The dependence of $\sigma_{-1n}$ on single-particle energy of the last occupied nucleon is seen in our present calculations. That means, although
the total nuclear reaction cross-section does not show a significant difference, the $\sigma_{-1n}$
values differ from each other depending on the NL3 or G2 parameter set.  The $\sigma_{-1n}$
are in good agreement with the experiments, when we consider the deformation effect in the densities.
The Glauber model fails for halo systems and in this case $^{19}$C+$^{12}$C is a
typical example. In such case, the difference between the total reaction
cross-section from the consecutive nuclei is applicable to evaluate $\sigma_{-1n}(II)$.
It is also concluded in the present paper
that the deformation effects for one neutron removal cross-section
is very much crucial unlike to the total reaction cross-section $\sigma_r$.
In other words, the Glauber model reproduce the experimental data pretty well while
considering the deformed densities for stable nuclei as projectile.  On the other
hand, when we estimate the difference
of reaction cross-section of nuclei with mass number A and that of A-1 in an
isotopic chain, we get good agreement with the experimental data for halo cases.

\section{Acknowledgments}

We are thankful to Dr. BirBikram Singh and Mr. M. Bhuyan for the fruitful discussions. 
This work is supported in part by
Council of Scientific $\&$ Industrial
Research (No.03 (1060) 06/EMR-II), as well as the
Department of Science and Technology, Govt. of India, project No. SR/S2/HEP-16/2005.

\end{document}